\documentclass[a4paper,11pt]{article}
\usepackage[total={5.8in, 8in}]{geometry}
\usepackage[utf8]{inputenc}
\usepackage{amssymb}
\usepackage{amsmath}
\usepackage{braket}
\usepackage{hyperref}
\usepackage{authblk}
\usepackage{graphicx}
\usepackage{cite}
\usepackage[utf8]{inputenc} 
\usepackage{subfig}
\usepackage{color}

\title{The LHC upper bounds for $pp \to \text{diboson}$, $t\bar t$ cross section 
on fermionic dark matter}

\author{Karim Ghorbani%
  \thanks{\texttt{karim1.ghorbani@gmail.com}}}
\affil{\it Physics Department, Faculty of Sciences, Arak University, Arak 38156-8-8349, Iran}

\author{Parsa Hossein Ghorbani%
  \thanks{\texttt{parsaghorbani@gmail.com}}}
\affil{\it Institute for Research in Fundamental Sciences (IPM)\\ 
 \it School of Particles and Accelerators, P.O. Box 19395-5531, Tehran, Iran}

\date{}

\begin{document}

\maketitle

\begin{abstract}

The ATLAS report in August 2016 provided an upper limit for the $pp\to$ diboson and $t\bar t$ cross sections. 
We consider a pseudoscalar-mediated fermionic dark matter together 
with gluon and photon effective operators interacting with 
the pseudoscalar. Choosing the resonance mass being $m_\rho=200, 750$ GeV and $2$ TeV, 
beside the relic density and the invisible Higgs decay constraints we 
constrain more the space of parameters with the diboson and $t\bar t$ cross section upper bounds.
We finally provide some benchmarks consistent with all the constraints. 
Having exploited a pseudoscalar mediator, the DM-nucleon cross section is velocity 
suppressed so that the model evades
easily the bounds put by the future direct detection experiments such as XENON1T. 

\end{abstract}
\newpage
\tableofcontents

\section{Introduction}
Last year in the early LHC Run 2 data with center-of-mass energy $13$ TeV, 
a seemingly excess in the diphoton events with the invariant mass of
about $750$ GeV and a best-fit width of about $45$ GeV was announced by ATLAS with 
local significances of $3.9 \sigma$ \cite{ATLAS}. 
The CMS collaboration had also reported 
the excess \cite{CMS:2015dxe} at diphoton invariant mass of about $760$ GeV, where the best-fit 
gave a narrow width and a local significance of $2.6\sigma$. 
If such an excess existed, similar to the discovery of the Higgs particle in 2012 \cite{Aad:2012tfa},
the new particle could be a spin-even field i.e., a spin-$0$ or a spin-$2$ (graviton)
according to the Landau-Yang theorem.
Unfortunately a next report by ATLAS \cite{ATLAS:2016eeo} announced that the excess disappeared after analyzing
more data. If it was not merely a statistical fluctuation then 
the first hint into beyond the standard model (BSM) had been found.

The ATLAS report \cite{ATLAS:2016eeo} however still provides precious upper bounds for diboson and $t\bar t$ final 
states cross sections. 
In this paper we investigate if a new resonance with mass $m_\rho$ shows up at the LHC, 
assuming that the resonance comes from a pseudoscalar particle decaying into diboson and $t \bar t$, 
how the fermionic dark matter scenario fits with the LHC upper bounds announced recently.

In the ATLAS report \cite{ATLAS:2016eeo}, the total decay width over the resonance mass, $m_\rho$,  has been taken within 
$\Gamma^{\text{tot}}/m_\rho=0.02-0.1$, while the resonance mass varies from $200$ GeV to $6$ TeV. 
In our computations we take the total decay width ratio to be $\Gamma^{\text{tot}}/m_\rho=0.03-0.06$ and pick three 
samples of the resonance mass being $m_\rho=200, 750$ GeV and $2$ TeV only for illustration.

We study the case where the new resonance is a pseudoscalar
and the dark matter candidate is a singlet Dirac fermion 
(see \cite{Fedderke:2014wda,Cirelli:2005uq,Baek:2011aa,Ghorbani:2014qpa} for examples on fermionic DM). 
The pseudoscalar in this model beside interacting with the Dirac fermion dark matter and the standard model Higgs couples also
to the gluons and the photon through the effective operators of dimension five we introduce in the model. The effective 
couplings then are bounded by the cross section upper limits at the LHC. 

A special feature of having a pseudoscalar mediator in the current model is that the DM-nucleon elastic scattering 
cross section is velocity suppressed and the model evades easily the constraint from direct detection experiments 
like LUX and XENON100 or even XENON1T. 

The paper is written with the following parts. In the next section we introduce the dark
matter model which possesses a pseudoscalar mediator and a fermionic DM candidate.
Then in section \ref{width} we study the necessary decay widths we use in our analyses. 
Available constraints on the model parameter space are discussed in section \ref{existing-constraints}. 
In section \ref{results} we show that dark matter masses even outside
the resonance region is consistence with the decay width of $25-45$ GeV and in the subsequent section
the upper bounds on the $pp$ cross sections are applied. 
We conclude the paper in section \ref{conclude}.

\section{Pseudoscalar Mediator}\label{model}

In this section we introduce our model against which we will examine the diboson and $t\bar t$
cross section amplitude bounds obtained in the ATLAS/CMS experiments. The model includes a Dirac fermion
dark matter candidate and a pseudoscalar together with two effective operators which are the sources for some of the 
processes measured at the LHC we analyze more in section \ref{dicross}.  

The pseudoscalar plays the role of a mediator between the dark sector and the SM sector. 
We suppose that the pseudoscalar field couples to the SM fields through
a gluon and a photon dimension 5 effective operators and a Higgs portal.
In the effective operators the pseudoscalar is coupled to gluons and photons with 
dimensionful couplings at some scale $\Lambda$ that we fix it latter.
The dark sector Lagrangian for such a setting reads, 
\begin{equation}\label{lagDM}
 {\mathcal L}_{\text{Dark}} =  \bar{\chi} (i {\not}\partial-m_{\text{DM}}) \chi  
  +\frac{1}{2} \partial_{\mu} \phi \partial^\mu \phi 
  - \frac{m^{2}}{2}\phi^2 -\frac{\lambda_\phi}{4}\phi^4\,,
\end{equation}
where $\phi$ stands for the pseudoscalar and $\chi$ 
is the singlet Dirac fermion representing the dark matter candidate. 
The Lagrangian for the interactions is 
\begin{equation}\label{lagint2}
\begin{split}
\mathcal L_{\text{int}}   =  -ig_\chi\phi \bar{\chi}\gamma^{5}\chi 
- g_{H} \phi^2  H^{\dagger}H  
+ c_{g} \frac{\alpha_{s}}{\pi v_{H}} \phi G_{\mu\nu}\tilde{G}^{\mu\nu}
+ c_{\gamma} \frac{\alpha_{em}}{\pi v_{H}} \phi F_{\mu\nu}\tilde{F}^{\mu\nu}\,,
\end{split}
\end{equation}
where $G_{\mu\nu}$ and $F_{\mu\nu}$ are the colored $SU(3)_c$ and the electromagnetic $U(1)$ field strengths 
in the SM respectively.
The tilde denotes the dual of the field strength, e.g.,
$\tilde{G}^{\mu\nu}=\frac{1}{2}\epsilon^{\mu\nu\rho\sigma} G_{\rho\sigma}$. 
Having in mind that $\phi$, $\bar{\chi}\gamma^{5}\chi$, $\tilde{G}^{\mu\nu}$ 
and $\tilde{F}^{\mu\nu}$ are odd under CP transformation 
and $H$, $G^{\mu\nu}$ and $F^{\mu\nu}$ are CP even, the Lagrangians (\ref{lagDM}) and 
(\ref{lagint2}) 
are CP invariant. Lagrangian (\ref{lagint2}) incorporate a 
pseudoscalar-Higgs quadratic interaction term. We will study these two cases separately in the 
following sections. Moreover, the Higgs 
potential in the SM sector reads,
\begin{equation}
 V= \mu  H^\dagger H  + \lambda_H \left( H^\dagger H \right)^2.
\end{equation}

It is worth mentioning that in the model described above, it is assumed that the pseudoscalar has a Yukawa coupling, $y_{x}$,
to a vector-like exotic quark, $q_{x}$, in the fundamental representation of $SU(3)_{c}$ 
with the Lagrangian $\mathcal L_{\text{int}} \sim - y_{x} \phi q_{x}\gamma^{5} q_{x}$. 
The effective couplings, $c_{g}$ and $c_{\gamma}$ are then generated by integrating out the vector-like quark $q_{x}$.  

Moreover, the pseudoscalar couples to the SM quarks only via mixing with the SM Higgs. The coupling to the light quarks
are negligible and therefore the pseudoscalar production at the LHC is dominated by the gluon fusion.

Note that even though we have not included the effective operators such as 
$\phi W_{\mu\nu}\tilde{W}^{\mu\nu}$ and $\phi B_{\mu\nu}\tilde{B}^{\mu\nu}$ 
in the Lagrangian (\ref{lagint2}), however, we can implicitly have the 
pseudoscalar-gauge boson couplings through the mixing of the pseudoscalar and the Higgs.
The vacuum expectation value of the pseudoscalar can take a non-zero value, 
$\braket{\phi}=v_\phi$. For the Higgs particle the LHC has already fixed the mass to be $m_H\sim 125$ GeV 
and the Higgs vacuum expectation value is known, $v_H=246$ GeV. Having chosen a non-zero {\it vev} for the 
pseudoscalar there is a mixing between the Higgs and the pseudoscalar. 
Expressing the Higgs and the pseudoscalar fields by fluctuations around their 
{\it vev}s as $\phi=v_\phi+\rho'$ and 
$H^\dagger=\frac{1}{\sqrt{2}}\left( \begin{matrix}  
            0 & v_H+h'
           \end{matrix} \right)$,
and after diagonalizing the 
mass matrix, the mass eigenvalues (eigenstates) are described in terms of the
$m_{h'}$ (field $h'$) and $m_{\rho'}$ (field $\rho'$) and the mixing angle $\theta$.  
The mixing therefore opens a channel through which the pseudoscalar can decay into SM particles.
Denoting the Higgs and the pseudoscalar mass eigenstates by $h$ and $\rho$ respectively, the mass 
eigenvalues are given as the following, 
\begin{equation}
\begin{split}
 m^{2}_{h} = \frac{m^{2}_{h'}+m^{2}_{\rho'}}{2}+ \frac{m^{2}_{h'}-m^{2}_{\rho'}}{2} \sqrt{1+y^2}\,, 
 m^{2}_{\rho} = \frac{m^{2}_{h'}+m^{2}_{\rho'}}{2}- \frac{m^{2}_{h'}-m^{2}_{\rho'}}{2} \sqrt{1+y^2}\,,
 \end{split}
\end{equation}
where,  
\begin{equation}
 \tan(2\theta) = y= \frac{2m^{2}_{h'\rho'}}{m^{2}_{h'}-m^{2}_{\rho'}}, \hspace{.75cm} 
 m^{2}_{h'\rho'}= 2 g_{H} v_H v_{\phi}, \hspace{.75cm}
 m^{2}_{h'}= 2 \lambda_{H} v_H^{2}, \hspace{.75cm}
 m^{2}_{\rho'}= 2 \lambda_\phi v^{2}_{\phi}.
\end{equation}
The mass eigenvalues now are taken to be the physical mass of the Higgs and 
the mass of some  would-be resonances,
i.e., $m_h\equiv m_H\sim125$ GeV, $m_\rho = 200,750,2000$ GeV respectively. 
The stability conditions put already some constraints 
on the couplings of the model which are $\lambda_\phi>0$, $\lambda_H>0$ 
and $\lambda_\phi \lambda_H > 6 g^2_H $ (if $g_H<0$). 

It is most convenient to write out the quartic couplings in terms of the physical masses of the scalars and the mixing angle 
in following way,
\begin{equation}
 \begin{split}
 \lambda_{H}  = \frac{m^{2}_{\rho} \sin^2 \theta +m^{2}_{h} \cos^2 \theta }{2v^{2}_{H}}\,, 
 \nonumber\\ 
 \lambda_{\phi}  = \frac{m^{2}_{\rho} \cos^2 \theta +m^{2}_{h} \sin^2 \theta }{v^{2}_{\phi}/3}\,, 
 \nonumber\\
 g_{H} = \frac{m^{2}_{\rho}-m^{2}_{h}}{4v_{H}v_{\phi}} \sin 2\theta.
  \end{split}
\end{equation}
Since $m_{h}$ and $v_{H}$ are known and in this work we will choose $m_{\rho} = 200,750,2000$ GeV, 
we then take the set $\{ \theta,v_{\phi}, g_{\chi},c_{g}, c_\gamma\}$ as free parameters.

\section{Partial Decay Widths}\label{width}

We calculate the relevant partial decay widths when the interaction 
Lagrangian consists of two effective operators together with a Higgs portal. 
In this case, the pseudoscalar mixes with the SM Higgs. Therefore, the pseudoscalar decay channels additionally 
incorporate all the decay modes of the SM Higgs multiplied by a factor depending on the mixing angle.
All possible pseudoscalar decay modes are $\rho \to \chi \chi$, $\gamma \gamma$, $g g$, $W^+ W^-$, $Z Z$, $Z \gamma$, 
$hh$, $f \bar f$, where fermions in the SM are denoted by $f$. 
The decay width of the pseudoscalar when decays into a pair of DM is
\begin{equation}\label{dark-decay}
\Gamma_{\chi} 
= \Gamma( \rho \to \bar \chi \chi ) = \frac{g_{\chi}^2 m_{\rho} \cos^2\theta}{8\pi} (1-\frac{4 m^{2}_{\text{DM}}}{m^{2}_{\rho}})^{1/2} \,,
\end{equation}
where $\theta$ is the mixing angle defined in the previous section. 

Let us now consider the decay of a pseudoscalar to $\gamma \gamma$ and $gg$. 
Due to the mixing with the SM Higgs, the pseudoscalar decay into two photons occurs not only through contact interaction 
but also can occur through loop processes induced predominantly via $W^\pm$ bosons 
and heavy fermions, in particular the top quark. Taking into account both effects, the resulting decay width reads

\begin{equation}
 \Gamma_{\gamma} = \Gamma(\rho \to \gamma \gamma) = 
 (\frac{\alpha_{\text{em}}}{4\pi})^2 \frac{m_{\rho}^3}{16\pi v_{H}^2} |{\cal F}|^2\,,
\end{equation}
where, 

\begin{equation}
 {\cal F} = 
 {\cal F}_{W}(\beta_{W}) \sin \theta + \sum\limits_{f} N_{c} Q_{f}^2 {\cal F}_{f}(\beta_{f}) \sin \theta 
 +  64~c_{\gamma}^2 \cos \theta  \,,
\end{equation}
and $\beta_{f,W} = \frac{4m_{f,W}^2}{m_{\rho}^2}$.  
The loop functions ${\cal F}_{W}$ and ${\cal F}_{f}$ are defined as
\begin{equation}
\begin{split}
\mathcal{F}_{W}(\beta) = 2 + 3 \beta = 3 \beta (2-\beta) f(\beta) \,, \\
\mathcal{F}_{f}(\beta) = -2\beta \Big( 1+(1-\beta)f(\beta) \Big) \,, \\
\end{split}
\end{equation}
where,
\begin{equation}
f(\beta) = -\frac{1}{4} \Big( \log(\frac{1-\sqrt{1+\beta}}{1-\sqrt{1-\beta}})+ i\pi  \Big)^2.
\end{equation}

The pseudoscalar decay into two gluons is possible via an contact operator and through loop 
processes induced predominantly by heavy quarks. The final result for the decay width is

\begin{equation}
 \Gamma_{g} = \Gamma(\rho \to g g) =  \frac{\alpha_s^2  m_{\rho}^3}{72\pi^3 v_{H}^2} |{\cal F}|^2\,,
\end{equation}          
where,  
\begin{equation}
 {\cal F} = \sum\limits_{q} {\cal F}_{q}(\beta_{q})  \sin \theta 
+ 144~c_{g}^2 \cos \theta 
\end{equation}
and
\begin{equation}
 {\cal F}_{q}(\beta) = \frac{3}{2} \beta (1+(1-\beta) f(\beta)).
\end{equation}   

For the rest of the decay modes we can apply the known formulas given for the relevant SM-Higgs decays
which are now scaled by $\sin^2 \theta$. The decay width for fermion emission is 

\begin{equation}
\Gamma_{f} = 
\Gamma( \rho \to f \bar f ) = \frac{N_c}{8\pi} \frac{m_f^2}{v_H^2} m_{\rho} \sin^2 \theta  (1-4 x_f^2)^{3/2} \,,
\end{equation}
where $x_f = m_f/m_\rho$ and we set the color factor $N_c = 1$ for leptons and $N_c = 3$ for quarks.

The pseudoscalar can decay into $W^\pm$ gauge bosons with the following decay width
\begin{equation}
\Gamma_{W} = 
\Gamma( \rho \to W^+ W^- ) = \frac{1}{16\pi} \frac{m_\rho^3}{v_H^2} \sin^2 \theta  \sqrt{1-4x_W^2} (1-4 x_W^2 + 12 x_W^4) \,,
\end{equation}
where $x_W = m_W/m_\rho$. The pseudoscalar can decay into $Z$ bosons with the decay rate 
\begin{equation}
\Gamma_{Z} = 
\Gamma( \rho \to Z Z ) = \frac{1}{32\pi} \frac{m_\rho^3}{v_H^2} \sin^2 \theta  \sqrt{1-4x_Z^2} (1-4 x_Z^2 + 12 x_Z^4) \,,
\label{gammaZZ}
\end{equation}
where $x_Z = m_Z/m_\rho$.
In our computations we apply the decay width $\Gamma_{Z\gamma} = \Gamma(\rho \to Z \gamma) \sim 10^{-3} \sin^2 \theta$ GeV
obtained from the exact formulas given in \cite{Djouadi:2005gi}.
Finally, we present the partial decay width of the pseudoscalar into a pair of SM-Higgs bosons as
\begin{equation}
\Gamma_{h} = 
\Gamma( \rho \to h h ) = \frac{\alpha^2}{8\pi m_{\rho}} \sqrt{1-\frac{4m_h^2}{m_\rho^2}} \,,
\end{equation}
where $\alpha = (2\cos \theta -6 \cos \theta \sin^2 \theta)g_H v_\phi + (4\sin \theta - 6 \sin^3 \theta) g_H v_H
        +\cos \theta \sin^2 \theta \lambda_\phi v_\phi - 6 (\cos^2 \theta \sin \theta) \lambda_H v_H$.

\section{The Constraints}
\label{existing-constraints}
In this section we discuss the LHC constraints, constraints from the oblique parameters and the 
observed relic density.

\subsection{Higgs Physics Constraints}
Two new decay channels for the SM Higgs boson will be possible in 
case $m_{\rho} < m_{h}/2$ and $m_{\text{DM}} < m_{h}/2$. 
In the present work where $m_{\rho}  = 200,750, 2000$ GeV, only the decay $h \to \chi \chi$ can happen 
for small enough DM mass. Invisible Higgs decay investigations at the LHC put 
an upper limit on the invisible branching ratio, $\text{Br}_{\text{inv}} \lesssim 0.24$ \cite{CMS-PAS-HIG-16-016}.
Applying this experimental bound we find,  
\begin{equation}
|g_{\chi} \tan \theta| < \frac{5.04~(\text{MeV})^{1/2}}{(m^2_{h} - 4 m^{2}_{\chi})^{1/4}} \,.
\end{equation}
This will restrict our model parameter space in the regions with $m_{\text{DM}} < m_{h}/2$.

On the other hand, an observable $\mu$ called signal strength which is measured by ATLAS and CMS 
has the Following definition,

\begin{equation}
\mu_{i}^{f} = \frac{\sigma_{i} \times \text{Br}^{f}}{(\sigma_{i} \times \text{Br}^{f})_{\text{SM}}} \,,
\end{equation}
where $\sigma_{i}$ is the Higgs production cross section 
via channel $i$ and $\text{Br}^{f}$ is the branching ratio of Higgs decaying into a final state $f$. 
Given various Higgs production and decay channels, the LHC best-fit result is $\mu = 1.09^{+0.11}_{-0.10}$ \cite{Khachatryan:2016vau}.
Due to the mixing between the SM Higgs and the singlet pseudoscalar in our model, $\sigma_{i}$ is 
scaled by a factor $\cos^2 \theta$ while $\text{Br}^{f}$ remains the same as its SM value. 
Therefore, an upper limit of $\theta \lesssim 0.12$ on the mixing angle is found at $1\sigma$ level \cite{Ghosh:2015apa}.

\subsection{Oblique Parameters}
For small mixing angle only the oblique parameter $T$ is relevant. Following the discussion in \cite{Barger:2007im},
in the present model the parameter $T$ is given by

\begin{eqnarray}
  T^{BSM} = -\Big(\frac{3}{16\pi s_{w}^2}\Big)
   \Big\{ \cos^2 \theta \Big[\frac{1}{c_{w}^2} (\frac{m_{h}^2}{m_{h}^2-m_{Z}^2}) \ln \frac{m_{h}^2}{m_{Z}^2} 
         - (\frac{m_{h}^2}{m_{h}^2-m_{W}^2}) \ln \frac{m_{h}^2}{m_{W}^2}   \Big]
  \nonumber\\&&\hspace{-10cm}
  + \sin^2 \theta \Big[\frac{1}{c_{w}^2} (\frac{m_{\rho}^2}{m_{\rho}^2-m_{Z}^2}) \ln \frac{m_{\rho}^2}{m_{Z}^2} 
         - (\frac{m_{\rho}^2}{m_{\rho}^2-m_{W}^2}) \ln \frac{m_{\rho}^2}{m_{W}^2}   \Big]  
  \Big\} \,.
\end{eqnarray}
The quantity $T$ is obtained in the SM by setting $\theta = 0$. The constraint from electroweak fit 
is given for $\Delta T  = T^{BSM} - T^{SM}$ in \cite{Agashe:2014kda} as $\Delta T = 0.01 \pm 0.12$. 
The oblique parameter puts insignificant constraint for small mixing angle of size $\theta \lesssim 0.1$.

\subsection{Relic Density}

In two experiments by Planck and WMAP, the relic density of DM is obtained. The combined result 
is $0.1172 <  \Omega_{\text{DM}} h^2 < 0.1226$ \cite{Ade:2013zuv,Hinshaw:2012aka}. An updated value for the 
relic density can be found in \cite{Ade:2015xua}. 
We will use this result to constrain the model parameter space. To this end, we need to solve numerically the Boltzmann equation, 

\begin{equation}
 \frac{dn_{\chi}}{dt} = -3Hn_{\chi} - \langle \sigma_{\text{ann}}v_{\text{rel}} \rangle [n^{2}_{\chi}-(n^{\text{EQ}}_{\chi})^2 ] \,,
\end{equation}
which provides us with the time evolution of DM number density and hence 
the present value of the density as a function 
of the thermal averaged annihilation cross section, $\langle \sigma_{\text{ann}} v_{\text{rel}} \rangle$.

To do the DM phenomenology we implement our model into the program MicrOMEGAs \cite{Belanger:2013oya}.
This package in turn employs the program CalcHEP \cite{Belyaev:2012qa} to compute the annihilation cross sections.

\begin{figure}
\begin{minipage}{0.51\textwidth}
\includegraphics[width=\textwidth,angle =0]{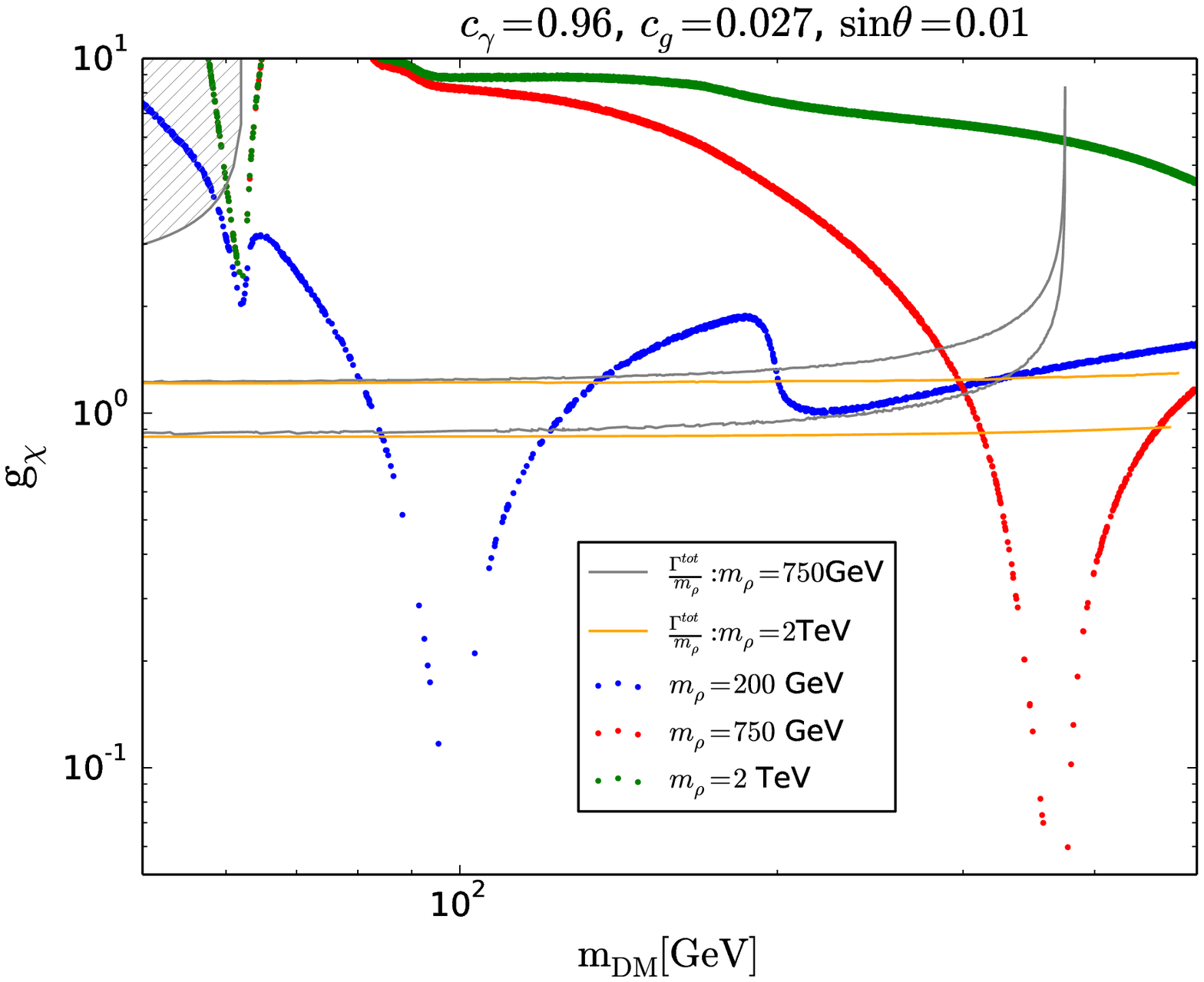}
\end{minipage}
\hspace{-.5cm}
\begin{minipage}{0.51\textwidth}
\includegraphics[width=\textwidth,angle =0]{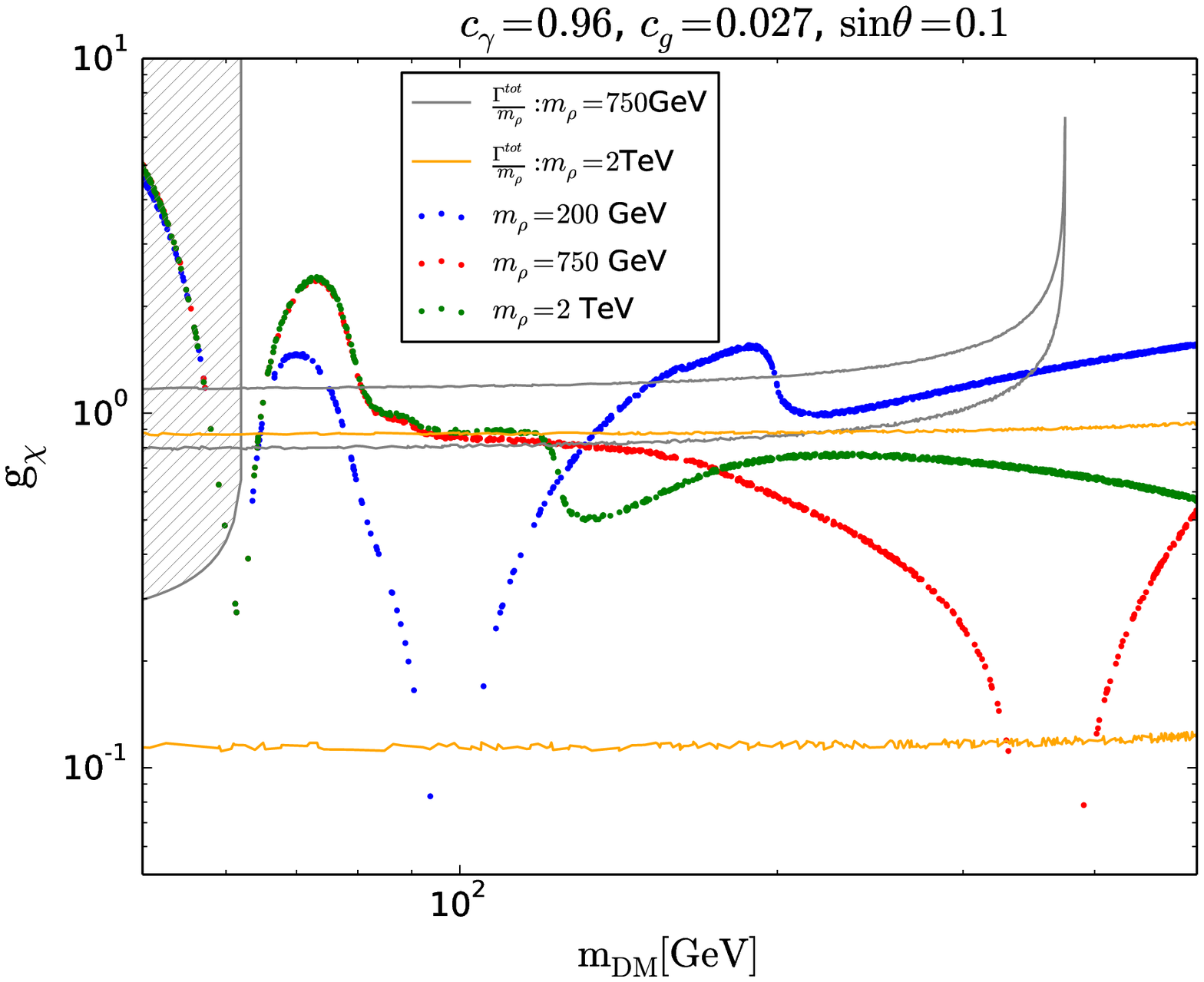}
\end{minipage}
\begin{minipage}{0.51\textwidth}
\includegraphics[width=\textwidth,angle =0]{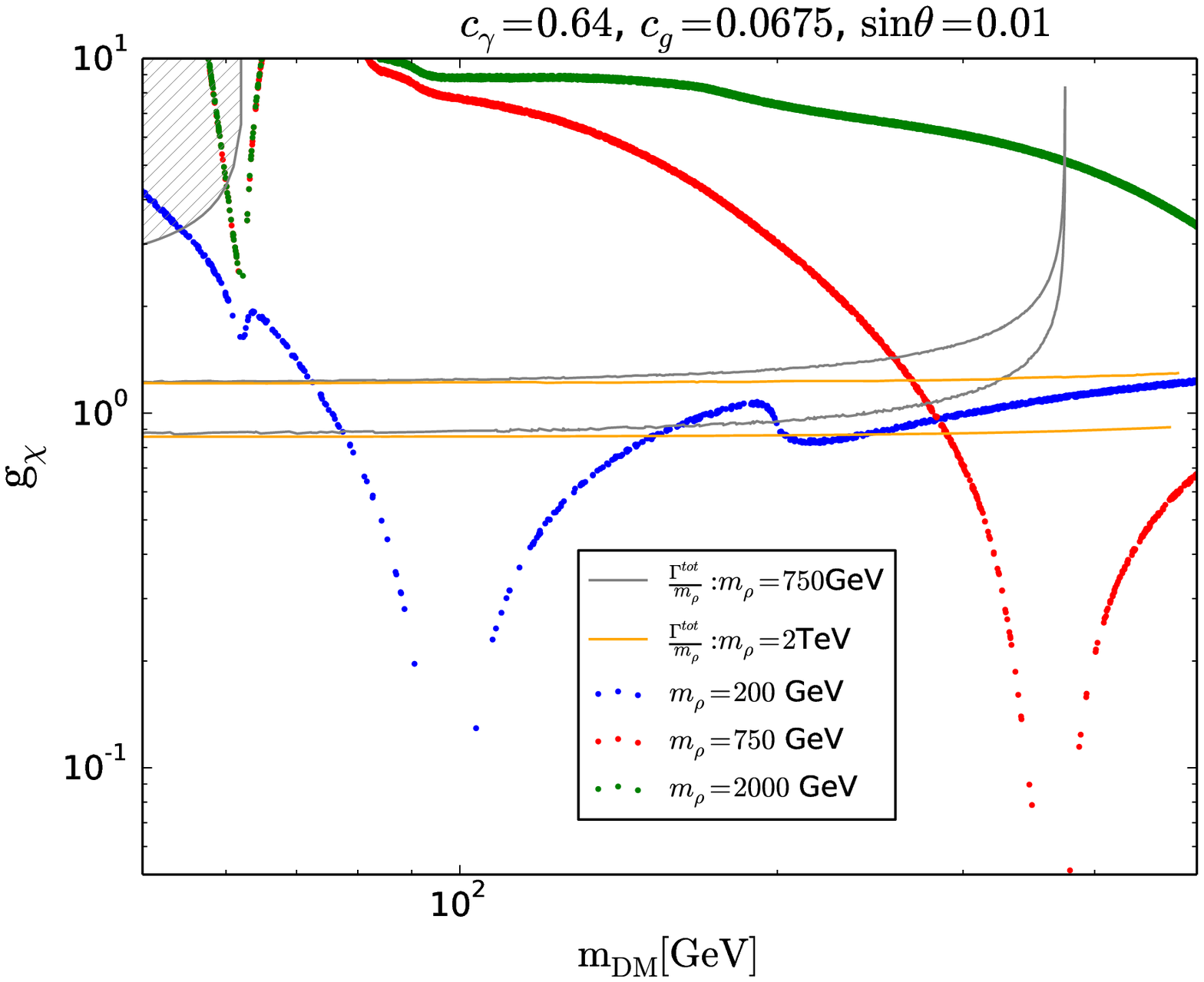}
\end{minipage}
\begin{minipage}{0.51\textwidth}
\includegraphics[width=\textwidth,angle =0]{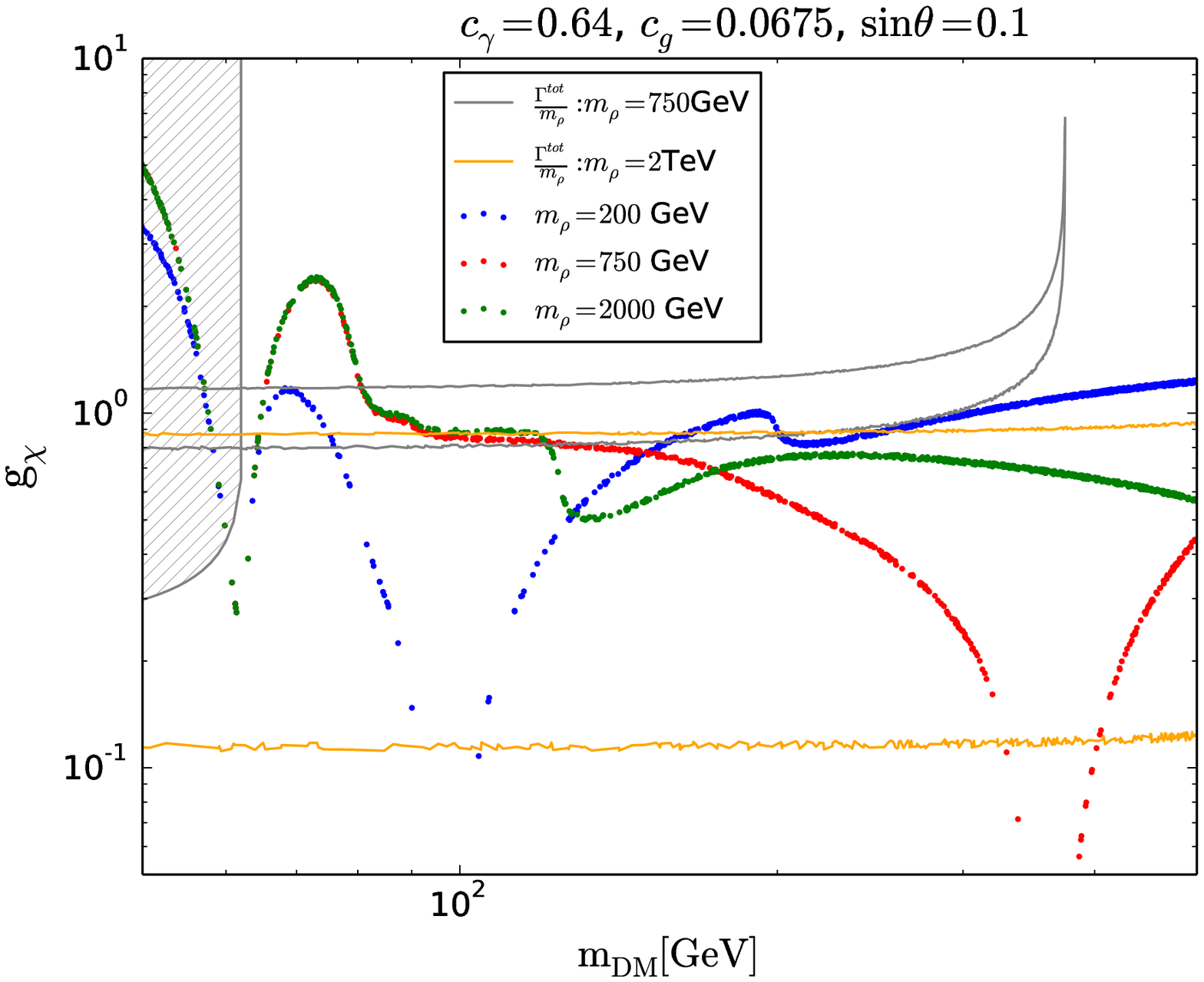}
\end{minipage}
\caption{The DM mass against the dark coupling is shown with the mediator mass $m_\rho=200, 750, 2000$ GeV respecting 
the WMAP/Planck relic density. The viable region is the intersection of the relic density line with the area 
of the total decay width
$\Gamma^{\text{tot}}/m_{\rho} \sim 0.03-0.06$. The plots are drawn for two sets of effective couplings 
$c_{\gamma} = 0.96, c_g=0.027$ and $c_{\gamma} = 0.64, c_g=0.0675$ and two mixing angles $\sin\theta =0.01, 0.1$.
The total decay width bound for $m_\rho=750$ GeV and $m_\rho=2000$ GeV  is the area between the gray lines and the 
orange lines, respectively. The shaded area is excluded by the invisible Higgs decay.}  
\label{relic-width1}
\end{figure}

\section{The Viable Parameter Space}\label{results}

In the present model any possible resonance is a pseudoscalar that plays the role of the mediator 
between the SM and the DM sectors.
Having introduced two effective operators of dimension five, the DM annihilation channels
are now $\chi \chi \to W^+ W^-, Z Z, h h, \bar f f, \gamma \gamma, gg$.
The SM fermions are denoted by $f$. One question that we would like to address here is 
whether there can be viable regions in the DM sector which is consistent both with the constraints coming 
from the resonance of the mass $\sim 200, 750, 2000$ GeV with the decay width 
in the range $\Gamma^{\text{tot}}/m_{\rho} \sim 0.03-0.06$, and constraints from observed relic density.  
We perform our computations for two sets of the effective couplings: $c_{\gamma} = 0.96, c_g=0.027$ and 
$c_{\gamma} = 0.64, c_g=0.0675$, with two values of the mixing angle, $\sin\theta =0.01, 0.1$. 
In all cases we choose $v_\phi=1000$ GeV.

Our numerical results for the two set of the effective couplings are 
shown in Fig.~\ref{relic-width1} for the DM mass being in the 
range $50$ GeV up to $500$ GeV, and for $\sin\theta= 0.01, 0.1$. 
It is evident from Fig.~\ref{relic-width1} that the role of the 
mixing angle is quite subtle in finding the DM mass range which gives both the relic density
and the anticipated resonance decay width correctly. 
Let us look at the results for the large mixing angle, i.e., $\sin \theta = 0.1$.
For the mediator mass, $m_{\rho} = 750$ GeV, 
there can be found DM candidates with mass $\sim 65$ GeV and $\sim 80-120$ GeV 
giving the observed relic density and the anticipated total decay width of the resonant. 
For the mediator mass, $m_{\rho} = 2000$ GeV, the viable region is $m_{\text{DM}}=65$ GeV and $m_{\text{DM}}>90$  GeV.
For the smaller resonance mass, $m_{\rho} = 200$ GeV, the total decay width does not sit in the range 
$\Gamma^{\text{tot}}/m_{\rho} \sim 0.03-0.06$
because the decay channels $\rho\to hh, t\bar t$ are no longer possible . 

It can be seen readily that our results do not change much by going from one set of the couplings 
$\{ c_{\gamma}, c_{g} \}$ to the other one.

\section{Effective couplings consistent with the LHC bounds and the DM constraints} \label{dicross}
\begin{table}
\centering
\begin{tabular}{p{7.5cm}|p{4.5cm}}
 \hline
 ~~~~~~~~~~~$\sqrt{s} = 13$ TeV   & ~~~~~~~~~~  $\sqrt{s} = 8$ TeV   \\ \hline

 $\sigma(pp \to W^+W^-) < 300$ fb   \cite{ATLAS-CONF-2015-075}     &    $\sigma(pp \to t\bar t) < 700$ fb      ~~\cite{Aad:2015fna}          \\ 
 $\sigma(pp \to ZZ) < 200$ fb      ~~~~~ \cite{ATLAS-CONF-2015-071}     &    $\sigma(pp \to gg) < 2.2$ pb ~\cite{CMS-PAS-EXO-14-005}    \\ 
 $\sigma(pp \to Z\gamma) < 28$ fb  ~~~~~~~ \cite{ATLAS-CONF-2016-010}     &                                                 \\ 
  $\sigma(pp \to hh) < 120$ fb    ~~~~~~ \cite{ATLAS-CONF-2016-017}     &                 \\      
   $\sigma(pp \to \gamma\gamma) < 3$  fb  ($m_\rho=750$ GeV) \cite{ATLAS:2016eeo}&   \\
    $\sigma(pp \to \gamma\gamma)<0.3$  fb  ($m_\rho=2000$ GeV)  \cite{ATLAS:2016eeo}&  \\
      \end{tabular}
 \caption{Upper limits on the $pp$ cross sections for various final states provided by LHC at $\sqrt{s}= 8, 13$ TeV.}
\label{LHCbounds}
\end{table}
Assuming that a pseudoscalar resonance is responsible for the production of diboson 
and $t\bar t$ at the LHC, beyond the relevant background processes within the SM, 
we compute the various cross sections 
in terms of the introduced effective couplings, $c_{\gamma}$ and $c_{g}$. 
To this end, we first implement our model into 
FeynRules \cite{Alloul:2013bka} and then into MadGraph5 \cite{Alwall:2014hca}.  
Four benchmark points, $c_{\gamma} = 0.96, c_g=0.027$ and $c_{\gamma} = 0.64, c_g=0.0675$  each with 
$m_{\text{DM}} = 100~\text{GeV}, m_{\rho} = 750$~ GeV and $m_{\text{DM}} = 200~\text{GeV}, m_{\rho} = 2$ TeV 
are picked which already respect the observed relic density, $0.1172 <  \Omega_{\text{DM}} h^2 < 0.1226$,
and the anticipated pseudoscalar total decay width, $\Gamma^{\text{tot}}/m_{\rho} \sim 0.03-0.06$.
We compare these four benchmark points against the observed upper limits on the 
cross sections, $pp \to W^{+}W^{-}, ZZ, \gamma\gamma, hh, gg, t\bar t$ at the LHC given in Table~\ref{LHCbounds}.

\begin{figure}
\begin{tabular}{ccc}
\hspace{-1.5cm}  \includegraphics[width=55mm]{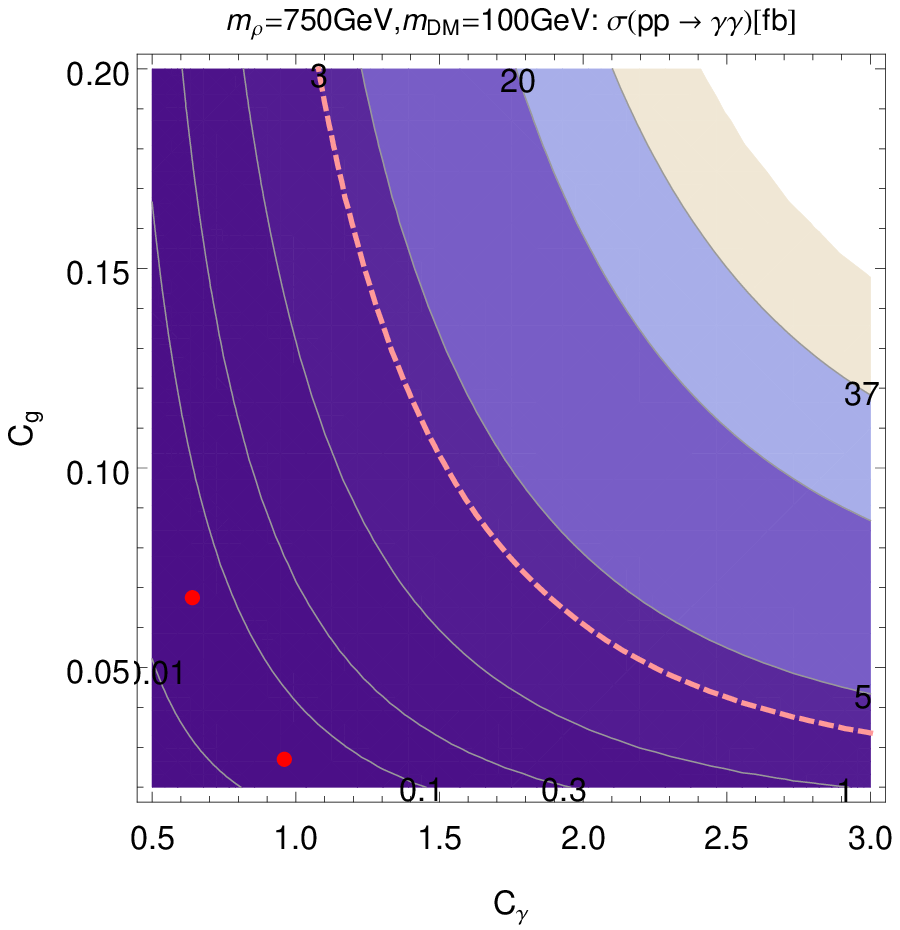} &   \includegraphics[width=55mm]{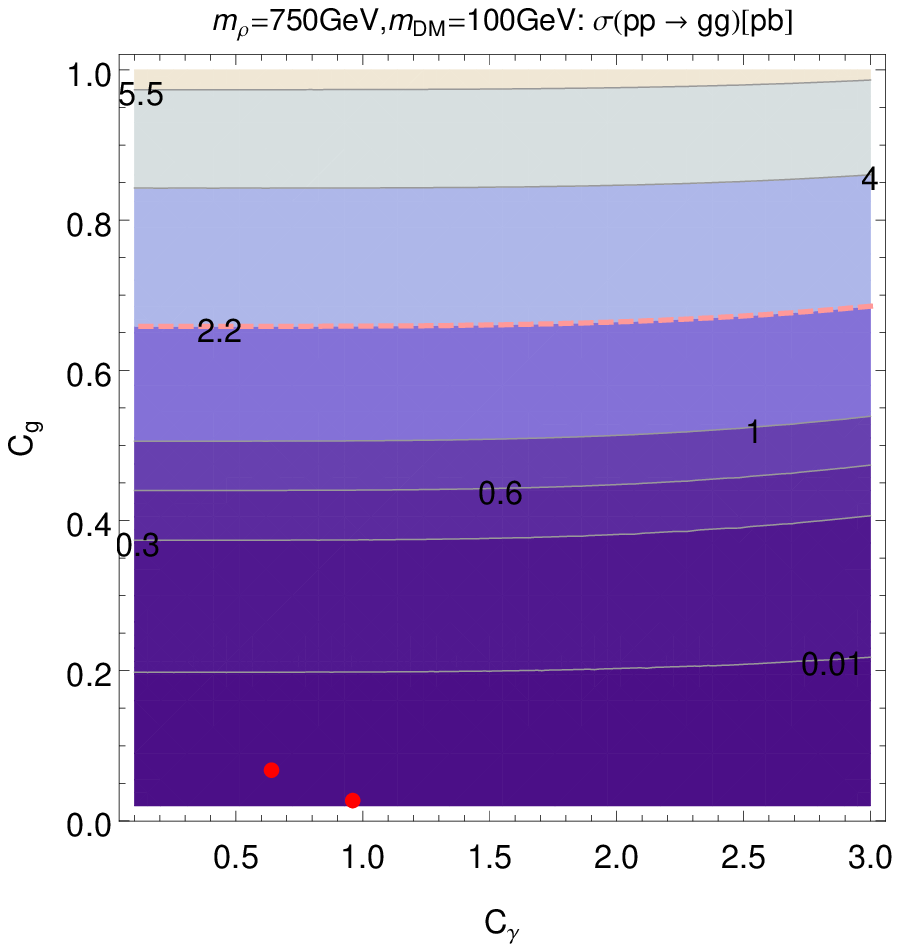} &   \includegraphics[width=55mm]{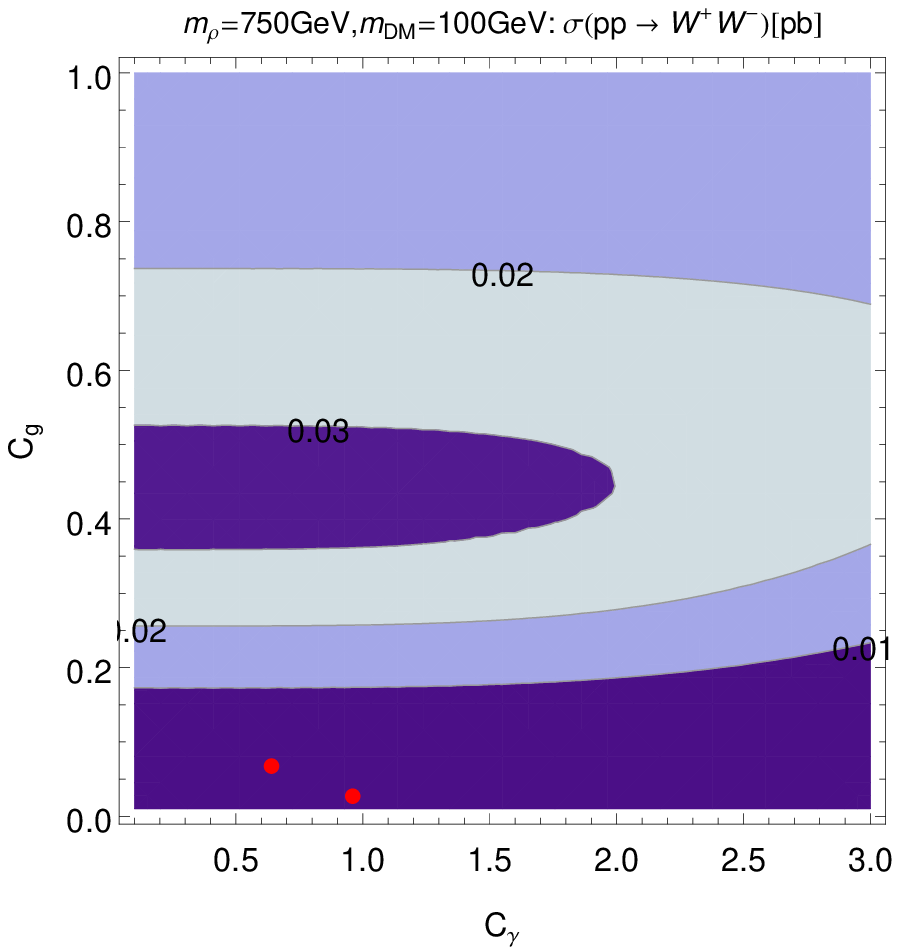} \\
 \hspace{-1.5cm}  \includegraphics[width=55mm]{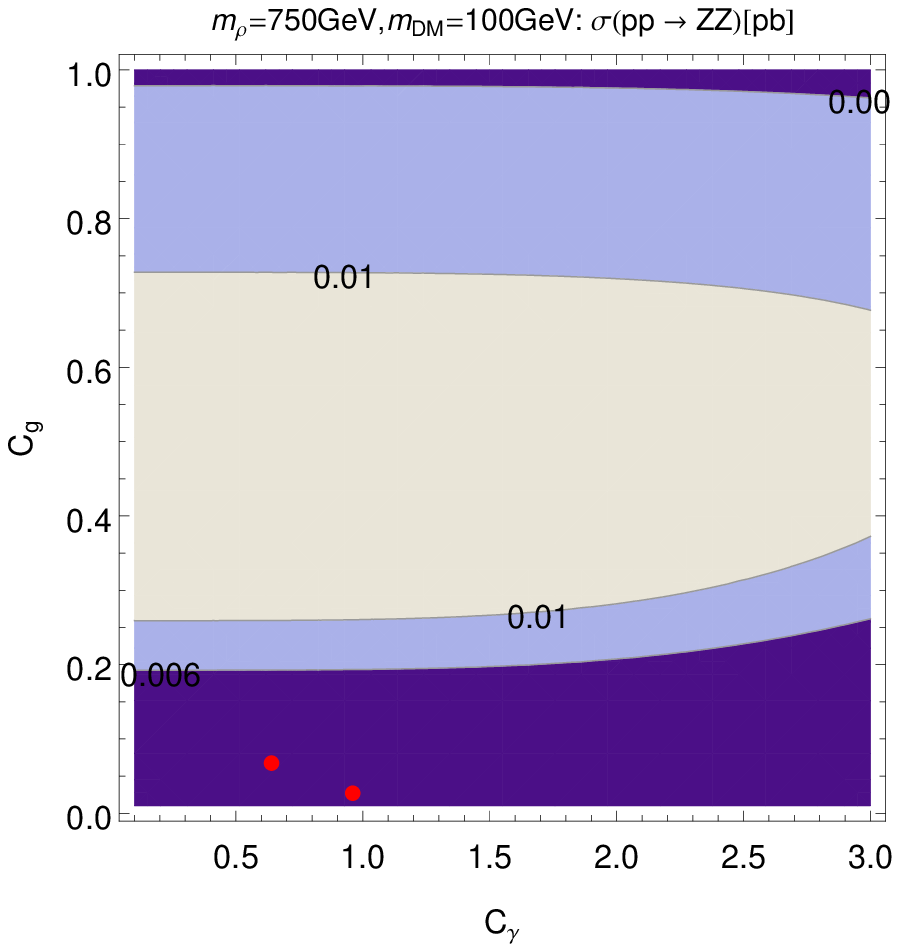}   &   \includegraphics[width=55mm]{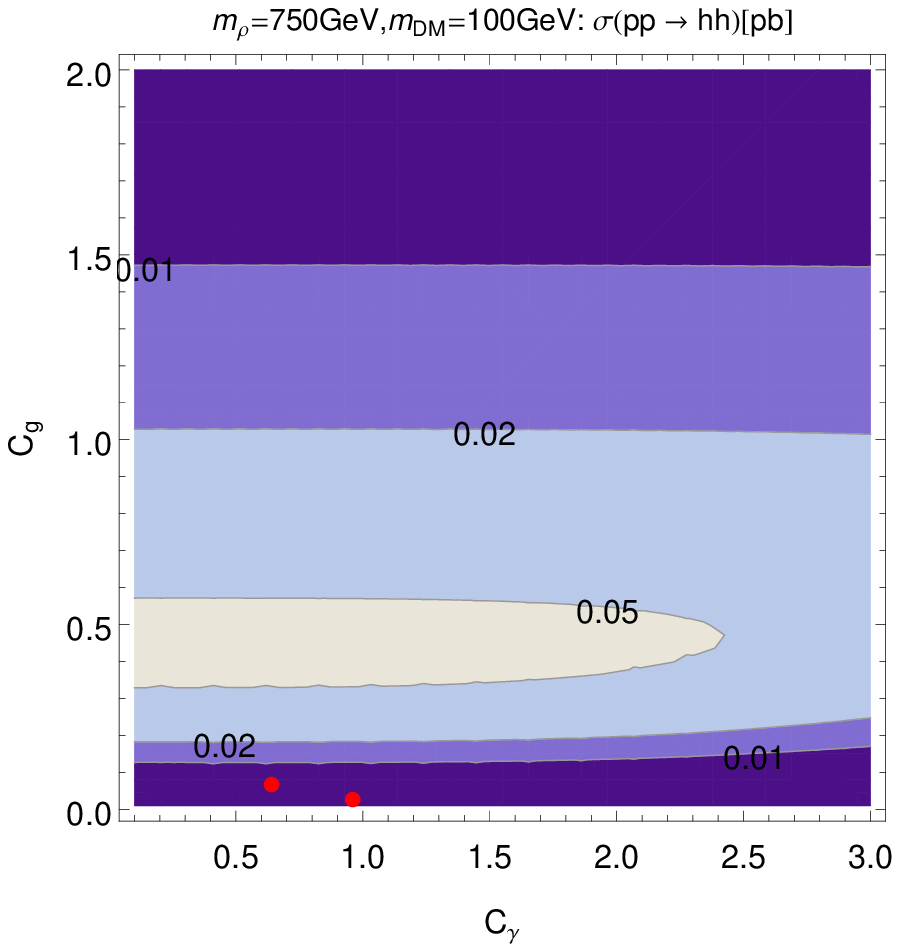} &  \includegraphics[width=55mm]{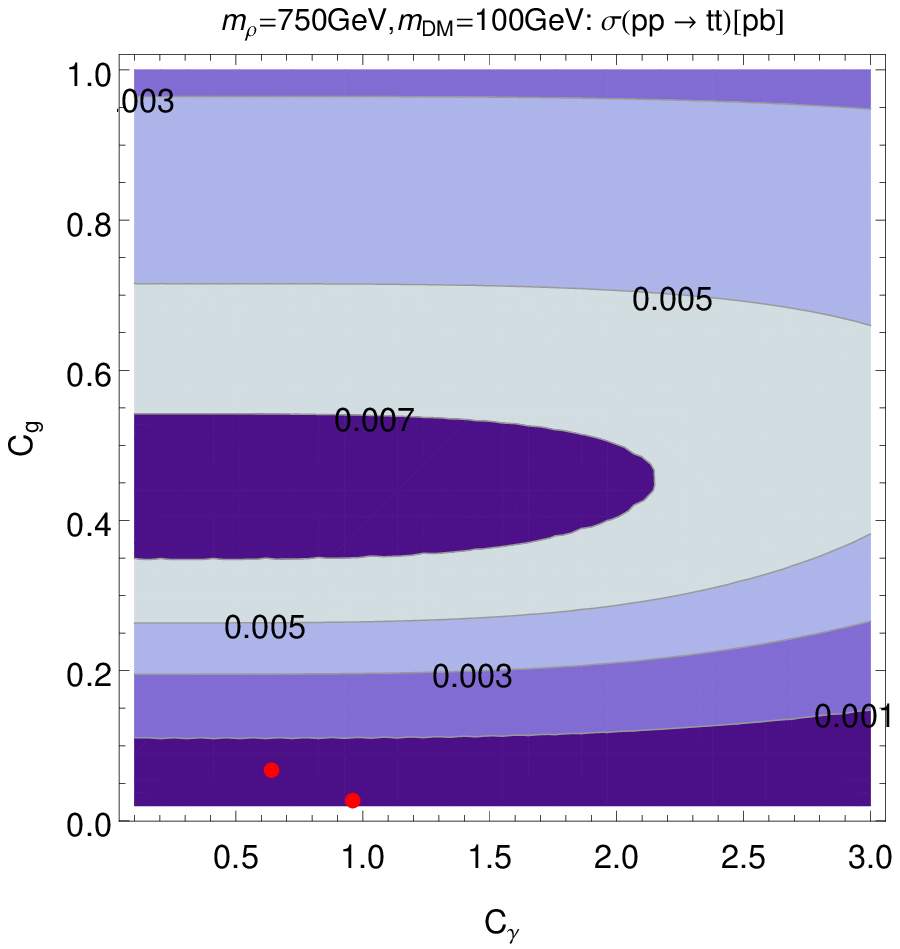} \\    
\end{tabular}
\caption{The contour plots illustrating the dependency of the various cross sections on the effective couplings with
$m_{\text{DM}}$=100 GeV and $m_\rho$=750 GeV. The dashed line shows the LHC upper bound on the cross section. The red points
are corresponding to two benchmarks $c_{\gamma} = 0.96, c_g=0.027$ and $c_{\gamma} = 0.64, c_g=0.0675$.} 
\label{benchmark1}
\end{figure}

\begin{figure}
\begin{tabular}{ccc}
\hspace{-1.5cm}  \includegraphics[width=55mm]{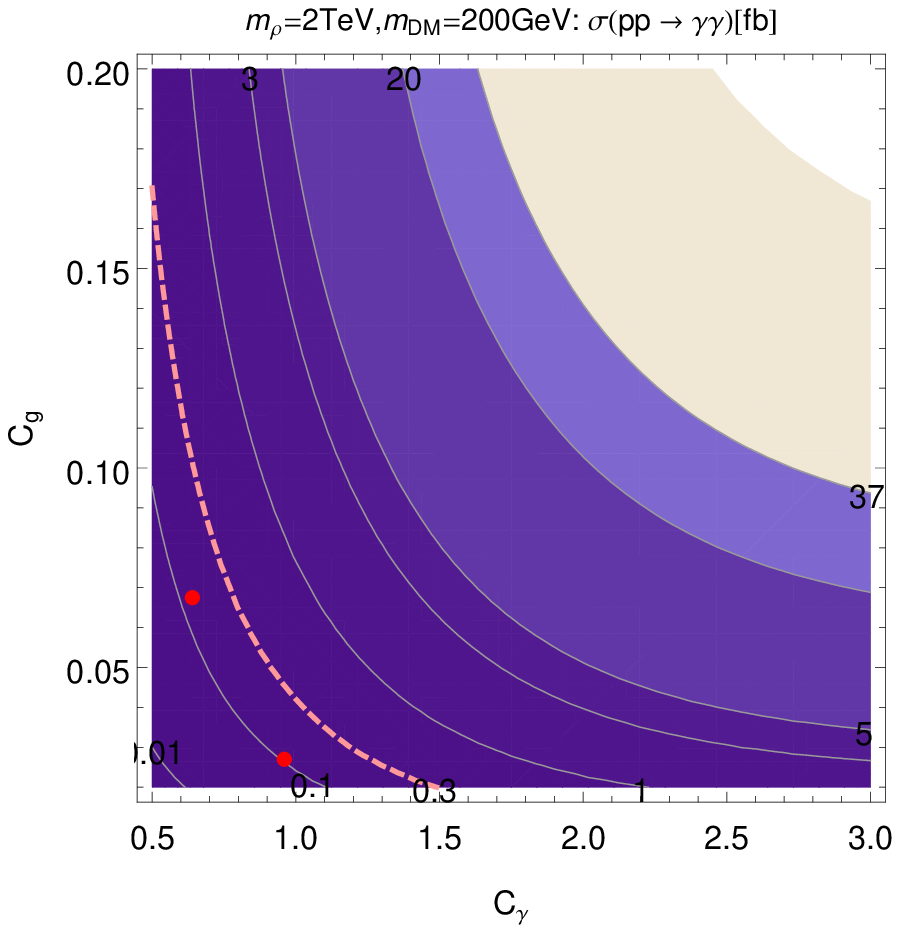} &   \includegraphics[width=55mm]{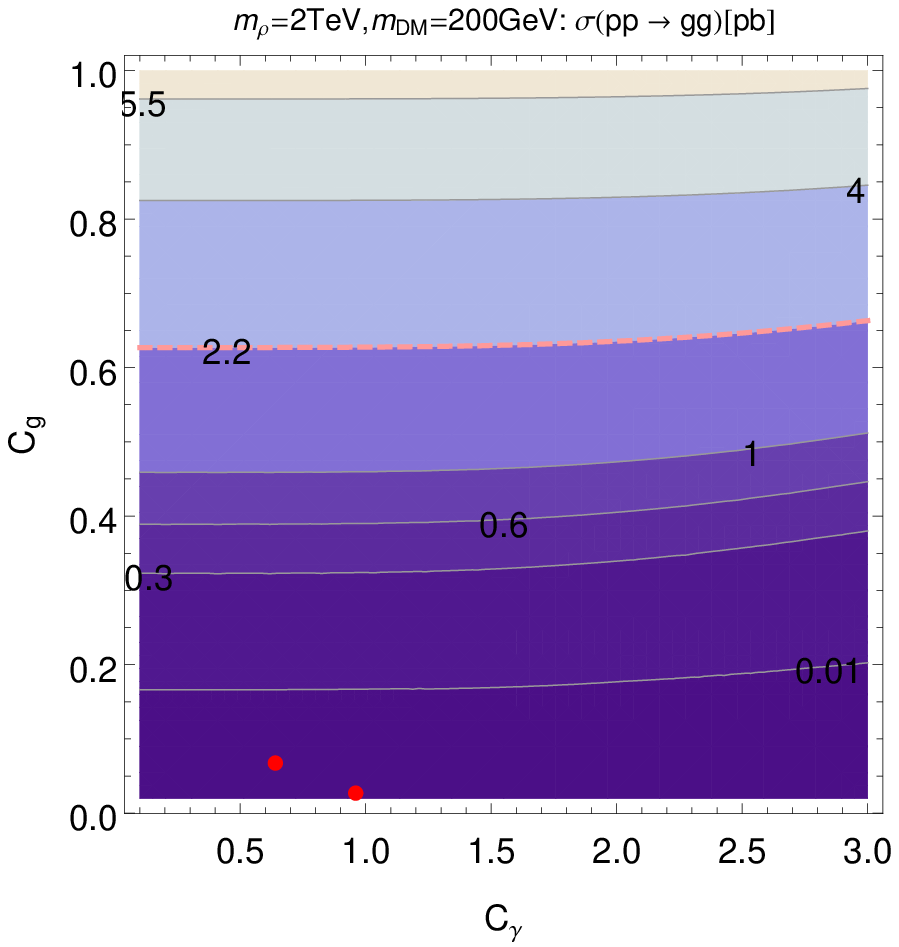} &   \includegraphics[width=55mm]{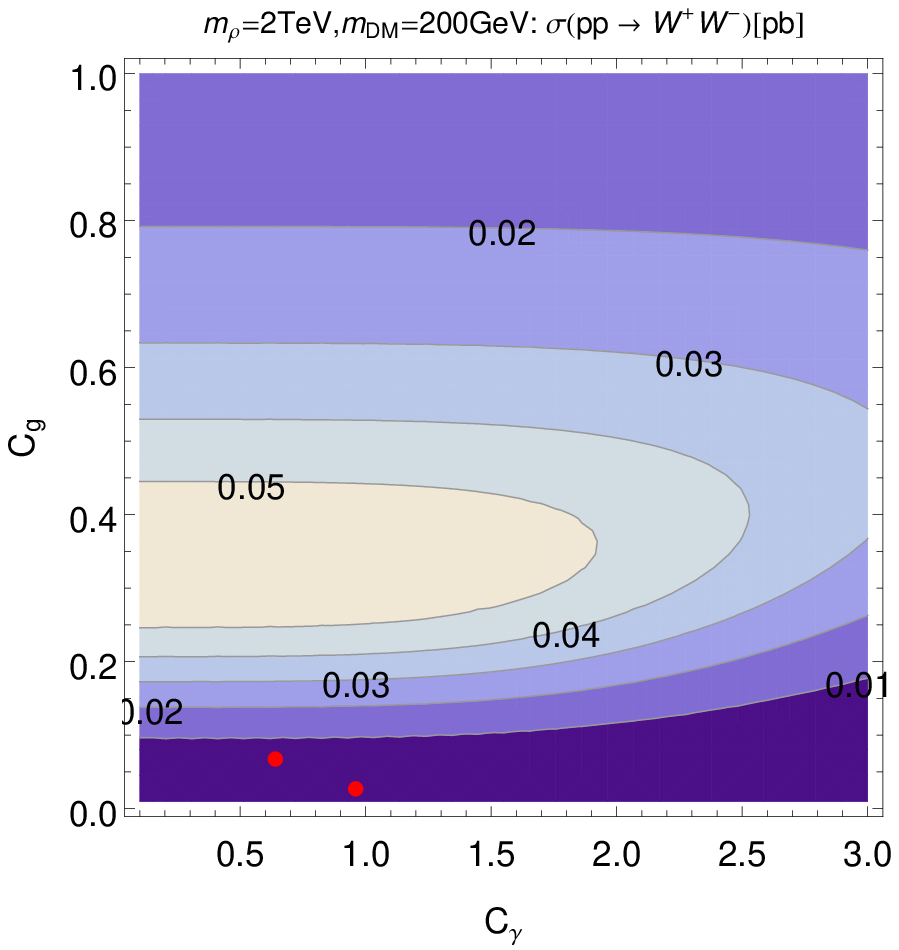} \\
 \hspace{-1.5cm}  \includegraphics[width=55mm]{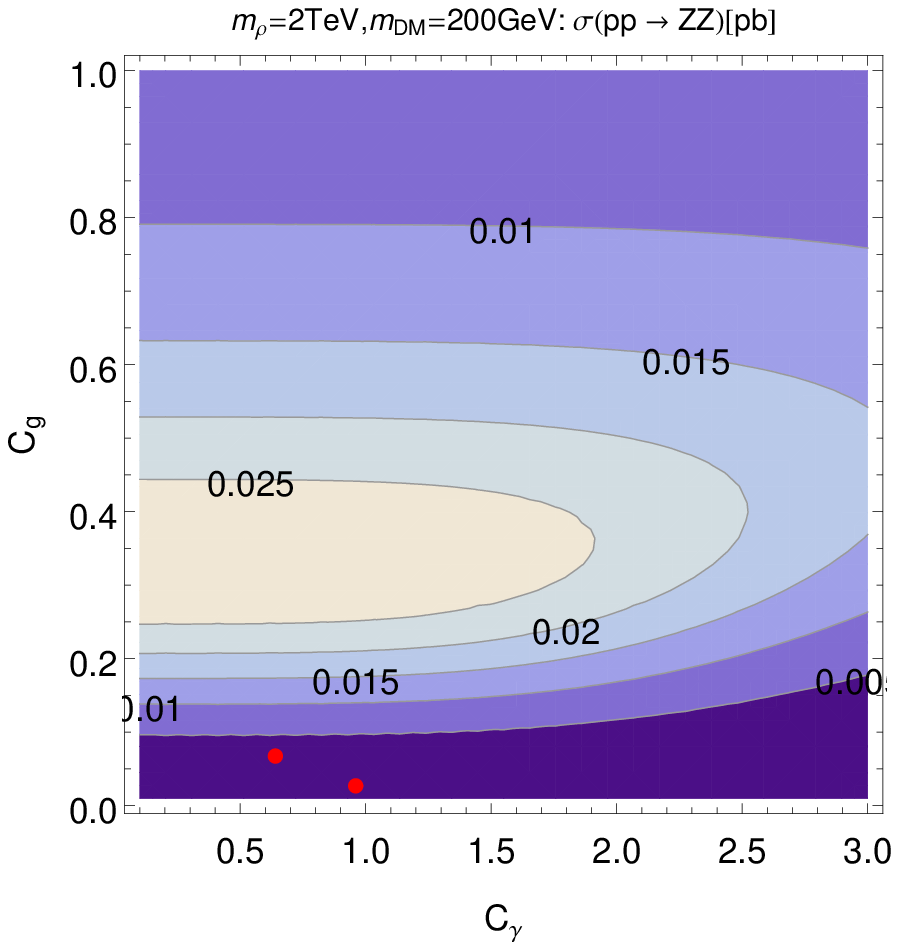}   &   \includegraphics[width=55mm]{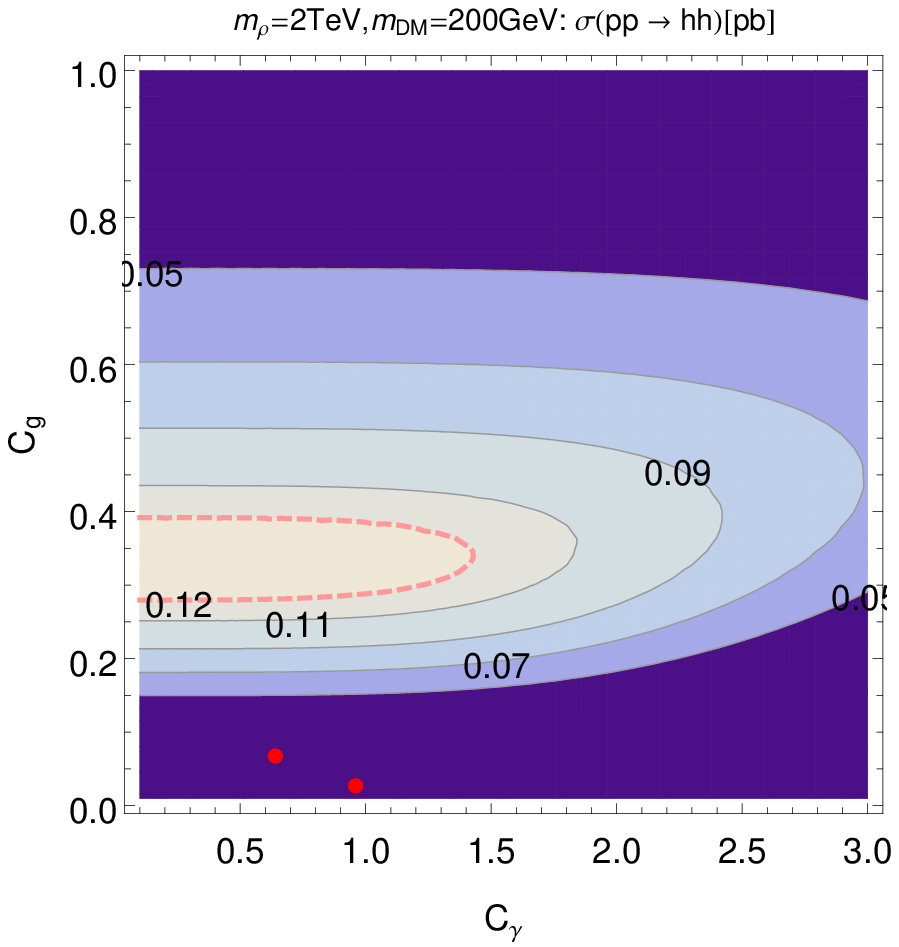} &  \includegraphics[width=55mm]{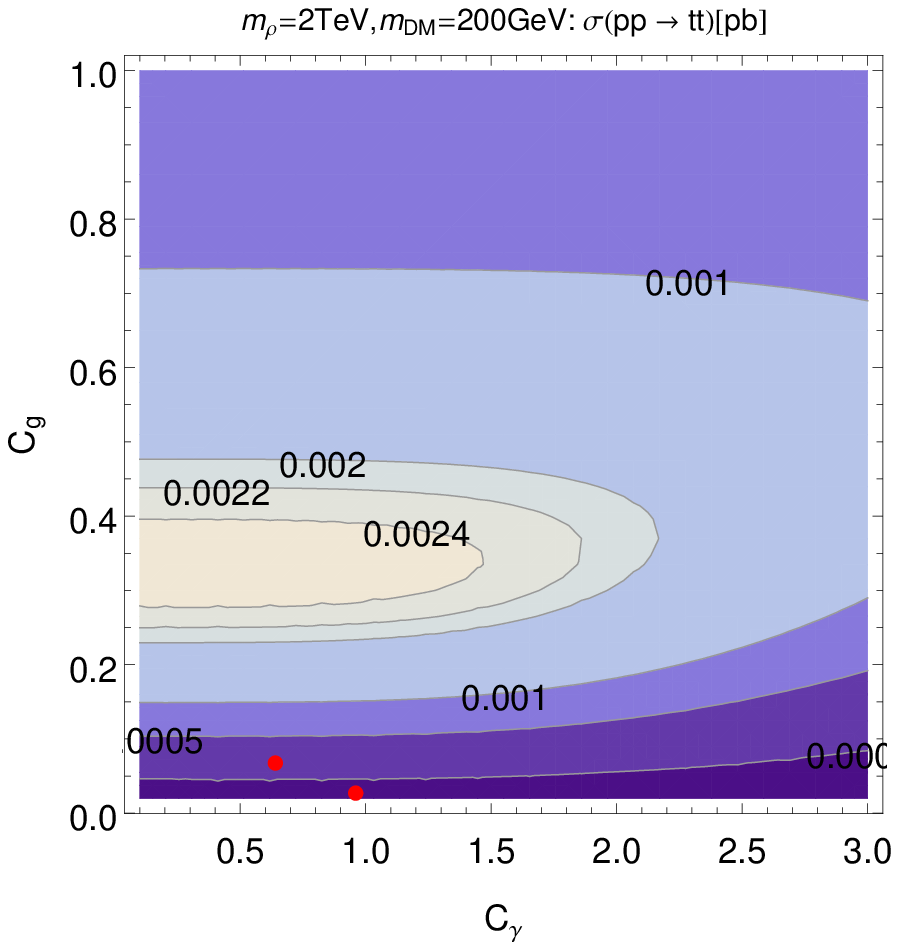} \\    
\end{tabular}
\caption{The contour plots illustrating the dependency of the various cross sections on the effective couplings with
$m_{\text{DM}}$=200 GeV and $m_\rho$=2000 GeV. The dashed line shows the LHC upper bound on the cross section. The red points
are corresponding to two benchmarks $c_{\gamma} = 0.96, c_g=0.027$ and $c_{\gamma} = 0.64, c_g=0.0675$.}
\label{benchmark2}
\end{figure}

In Fig.~\ref{benchmark1} and Fig.~\ref{benchmark2} we present our main results for the above mentioned cross sections
as contour plots and the corresponding upper limits  against the effective couplings 
for $m_{\text{DM}} = 100~\text{GeV}, m_{\rho} = 750$~ GeV 
and $m_{\text{DM}} = 200~\text{GeV}, m_{\rho} = 2$ TeV, respectively. 
In some plots the upper limits on the cross section reside beyond the 
range of the effective couplings or it is in a very small region and therefore are not visible.   
Strongest constraints come from the processes with $\gamma\gamma$ and $gg$ in the final 
state for $m_{\rho} = 750$~GeV and for $m_{\rho} = 2$~TeV with $\gamma\gamma$, $gg$ and $hh$ in the final state. 

Note that we have not included the process $pp \to Z\gamma$ in our plots. The reason is that 
$\Gamma(pp \to Z\gamma)$ is much smaller than $\Gamma(pp \to ZZ)$ as can be seen by comparing Eq.~\ref{gammaZZ} and the 
relation $\Gamma(\rho \to Z \gamma) \sim 10^{-3} \sin^2 \theta$ GeV. 
For instance when $m_{\rho} = 750$ GeV, $\Gamma_{ZZ}/\Gamma_{Z\gamma} \sim 280$. 
Given the upper limits for the two processes, $pp \to Z\gamma$ imposes 
much weaker constraints on the effective couplings.

\section{Conclusion}\label{conclude}

The exciting report by ATLAS and CMS in 2015 \cite{ATLAS,CMS:2015dxe} on a $750$ GeV excess in the 
diphoton events was nothing but a statistical fluctuation as announced by ATLAS 2016 report \cite{ATLAS:2016eeo}
and no significant excess was observed in 2016 data. Nevertheless the ATLAS 2016 report provided
an upper limit for the cross section of the 
diboson and $t\bar t$ in final state. 
In this paper we examined a fermionic dark matter scenario with 
a pseudoscalar mediator along with gluon and photon dimension five effective operators. 
The pseudoscalar plays the role of a spin-0 resonance which communicates with the standard model sector 
by the Higgs portal and couples also to the effective operators. 

We have taken three masses for the spin-0 resonance being $m_\rho=200, 750, 2000$ GeV and deal with two effective couplings 
$c_g$ and $c_\gamma$. 
In Fig. \ref{relic-width1} we have shown the viable DM mass for two sets of the effective couplings $\{c_g,c_\gamma\}$ 
and two mixing angles which fit with the observed relic density, invisible Higgs decay and gives the total 
decay width ratio $\Gamma^{\text{tot}}/m_\rho=0.03-0.06$. 

Then in contour plots Fig. \ref{benchmark1} and Fig. \ref{benchmark2} we have shown how the cross section 
for $pp\to W^+ W^-, ZZ, \gamma\gamma, gg, t\bar t$ depends on the 
effective couplings for $m_{\text{DM}}$=100 GeV, $m_\rho$=750 GeV and $m_{\text{DM}}$=200 GeV, $m_\rho$=2000 GeV respectively.
We have pinned down two benchmarks in each plot which fulfills all the constraints.

The characteristic of this fermionic dark matter model is that the DM-nucleon cross section 
is velocity suppressed because the mediator has been taken a pseudoscalar. The model therefore evades easily the 
bounds put by LUX \cite{Akerib:2016vxi} and XENON1T \cite{Aprile:2017iyp} or the future direct detection experiments.

\bibliography{ref}
\bibliographystyle{utphys}
\end{document}